\documentclass[runningheads]{utils/llncs}
\input{utils/collab}
\usepackage{cite}
\usepackage{amsmath,amssymb,amsfonts}
\usepackage{graphicx}
\usepackage{algorithmic}
\usepackage{graphicx}
\usepackage{textcomp}
\usepackage{xcolor}
\usepackage[strings]{underscore}
\usepackage{dblfloatfix}
\usepackage{tabularx}
\usepackage{booktabs}
\usepackage{paralist} 
\usepackage{xcolor}

\colorlet{punct}{red!60!black}
\definecolor{background}{HTML}{EEEEEE}
\definecolor{delim}{RGB}{20,105,176}
\colorlet{numb}{magenta!60!black}

\usepackage{color}

\definecolor{pblue}{rgb}{0.13,0.13,1}
\definecolor{pgreen}{rgb}{0,0.5,0}
\definecolor{pred}{rgb}{0.9,0,0}
\definecolor{pgrey}{rgb}{0.46,0.45,0.48}

\usepackage{listings}

\lstdefinelanguage{json}{
    basicstyle=\normalfont\ttfamily,
    numbers=left,
    numberstyle=\scriptsize,
    stepnumber=1,
    numbersep=8pt,
    showstringspaces=false,
    breaklines=true,
    frame=lines,
    xleftmargin=5.0ex,
    literate=
     *{0}{{{\color{numb}0}}}{1}
      {1}{{{\color{numb}1}}}{1}
      {2}{{{\color{numb}2}}}{1}
      {3}{{{\color{numb}3}}}{1}
      {4}{{{\color{numb}4}}}{1}
      {5}{{{\color{numb}5}}}{1}
      {6}{{{\color{numb}6}}}{1}
      {7}{{{\color{numb}7}}}{1}
      {8}{{{\color{numb}8}}}{1}
      {9}{{{\color{numb}9}}}{1}
      {:}{{{\color{punct}{:}}}}{1}
      {,}{{{\color{punct}{,}}}}{1}
      {\{}{{{\color{delim}{\{}}}}{1}
      {\}}{{{\color{delim}{\}}}}}{1}
      {[}{{{\color{delim}{[}}}}{1}
      {]}{{{\color{delim}{]}}}}{1},
}

\begin{document}
\title{Enabling Content Management Systems as an Information Source in Model-driven Projects}
\titlerunning{Enabling Content Management Systems as an Information Source}
%
\author{Joan Giner-Miguelez\inst{1}\orcidID{0000-0003-2335-6977} \and
Abel Gómez\inst{1}\orcidID{0000-0003-1344-8472} \and
Jordi Cabot\inst{2}\orcidID{0000-0003-2418-2489}}
%
%
\institute{Internet Interdisciplinary Institute (IN3), Universitat Oberta de Catalunya (UOC),  Barcelona, Spain \\
 \email{\{jginermi, agomezlla\}@uoc.edu}, 
\and
ICREA, Barcelona, Spain \\
 \email{jordi.cabot@icrea.cat}}
\maketitle              
\begin{abstract}
Content Management Systems (CMSs) are the most popular tool when it comes to create and publish content across the web. 
Recently, CMSs have evolved, becoming \emph{headless}.
Content served by a \emph{headless CMS} aims to be consumed by other applications and services through REST APIs rather than by human users through a web browser.
This evolution has enabled CMSs to become a notorious source of content to be used in a variety of contexts beyond pure web navigation. 
As such, CMS have become an important component of many information systems. 
Unfortunately, we still lack the tools to properly discover and manage the information stored in a CMS, often highly customized to the needs of a specific domain. Currently, this is mostly a time-consuming and error-prone manual process.

In this paper, we propose a model-based framework to facilitate the integration of headless CMSs in software development processes. Our framework is able to discover and explicitly represent the information schema behind the CMS. This facilitates designing the interaction between the CMS model and other components consuming that information. These interactions are then generated as part of a middleware library that offers platform-agnostic access to the CMS to all the client applications. The complete framework is open-source and available online.

\end{abstract}

\section{Introduction}


Content Management Systems (CMSs) are one of the most popular options to create content across the web. These systems, such as WordPress, Drupal, or Joomla, represent nearly 61,3\% in terms of published websites \cite{buildwith_2020,w3techs_2020}. One of the main reasons for this popularity is the great user experience provided by CMSs while creating content that empowers non-technical users to be part of the content creation chain \cite{cabot}. Besides, CMSs have evolved in the last years by implementing APIs to allow external apps to discover and interact with the content of these systems. This evolution, known as \emph{headless CMSs}, has shifted the main focus of these systems, traditionally on desktop solutions, to other kinds of applications such as mobile apps and other front-end apps.

As an example, in the media industry, CMSs are widely used to build digital solutions. Sony Pictures\footnote{\url{https://www.sonypictures.com/tv}}, Le Figaro\footnote{\url{https://www.lefigaro.fr}}, and Syfy Channel\footnote{\url{https://www.syfy.com}} are some of the media companies powering their digital solutions with open-source CMSs. The content created inside these solutions, such as news, podcast, or video, represents the key asset of these companies. Therefore, any new software development project in these companies will likely depend on, and require, the content in the CMSs.

This integration is becoming more difficult as CMSs solutions are becoming increasingly complex as so they do the APIs they expose. Moreover, typically, large companies must deal with various deployed CMSs, some of them legacy versions, that need to be combined to satisfy the application informational needs.  There is a clear need for new methods that help in managing these complex integration processes as part of new software development projects.
In parallel, there is increasing adoption of the Model-Driven Engineering (MDE) practices \cite{liev:tich:hans,huth:roun:stei} as MDE has been proven useful to tame the development complexity. Unfortunately, while we have several solutions to facilitate the integration of SQL \cite{egea2010mysql4ocl,Nguyen2019} and No-SQL \cite{Gwendal2019,abdelhedi2017} backends in MDE-based processes, there is a lack of solutions to support applications that rely on headless CMSs as a content source. Therefore, the integration of headless CMSs with other apps has been done by manual solutions, being these solutions time-consuming and error-prone.

In this work, we propose a framework to enable the integration of headless CMSs in MDE-based software development processes. The framework is composed of
\begin{inparaenum}[\itshape(i)\upshape]
\item a core model of CMSs,
\item a reverse-engineering process to extract the model from an existing deployed CMS in a UML representation, and finally,
\item a code-generation process that generates a middleware library to bridge the gap between the front-end consumer app and the CMSs.
\end{inparaenum} The framework focuses on existing CMSs rather than on creating new ones from a UML model because we have detected that most of the CMSs are already created, and there is a need to use them as a content source. 

With our framework, companies can quickly discover the information schema behind each CMS and represent it explicitly as a UML model. This model can then be integrated with other software models in the project. Finally, our framework also assists in the generation phase, simplifying the writing of all the required \textit{glue} code between the different components. As an outcome, we have built our proposal as an Eclipse plug-in and can be found in an open repository\footnote{\url{https://hdl.handle.net/20.500.12004/1/C/RCIS/2022/001}}.

The paper is structured as follows. Section \ref{sec:framework} presents the general overview of the framework. Section \ref{sec:CoreCMSModel} presents our \emph{Core CMS model} and, Section \ref{sec:reverse} presents the reverse engineering process. Section \ref{sec:integration} shows an example of integration and, Section \ref{sec:connector} presents the generation process. Finally, Section \ref{sec:tool} presents the developed tool, Section \ref{sec:relwork} summarizes the related work, and Section \ref{sec:conclusions} wraps up the conclusions and the future work.

\section{Framework overview}
\label{sec:framework}

In this section, we provide an overview of the proposed framework. The main goals of the framework are \begin{inparaenum}[\itshape(i)\upshape] 
\item to extract and explicitly represent the specific model of a deployed CMS to enable the integration of CMSs as information sources within a global model-based development process, and
\item to generate the \emph{glue code} that will transform any model-level interaction between the CMS model and other models (e.g. GUI models) in proper calls to the Headless CMS API. 
\end{inparaenum}

Following Fig.~\ref{fig:framework}, the first step is the \emph{Reverse engineering} process. This process takes as input the target CMS (in particular, its URL and user credentials), and imports the predefined \emph{Core CMS model} with basic CMS concepts. Then, creates the target CMS model as an extension to the core CMS one. The discovery process is implemented as a set of calls to the CMS API, adapted to the API offered by each CMS platform. We use UML to represent these models. 

As this obtained model is a purely standard model, designers can refine it and combine it with other models. For instance, we can easily model how a User Interface model queries the CMS model to retrieve the information it needs to show to the user in a platform-independent way.

The final step is the \emph{Code generation} process. This process takes as input the \emph{CMS Driver} that corresponds to the CMS platform and uses it to generate a middleware library allowing consumer apps to get the information they need from the CMS. Note that the API offered by the middleware only depends on the CMS content schema, not on the underlying CMS platform. The API is consistent across any CMS technology which facilitates any future CMS migration. As part of the API we have all the usual filtering, ordering and pagination methods that may be needed to iterate on the CMS content.

\begin{figure}[b]
    \centering
     \includegraphics[width=0.8\columnwidth]{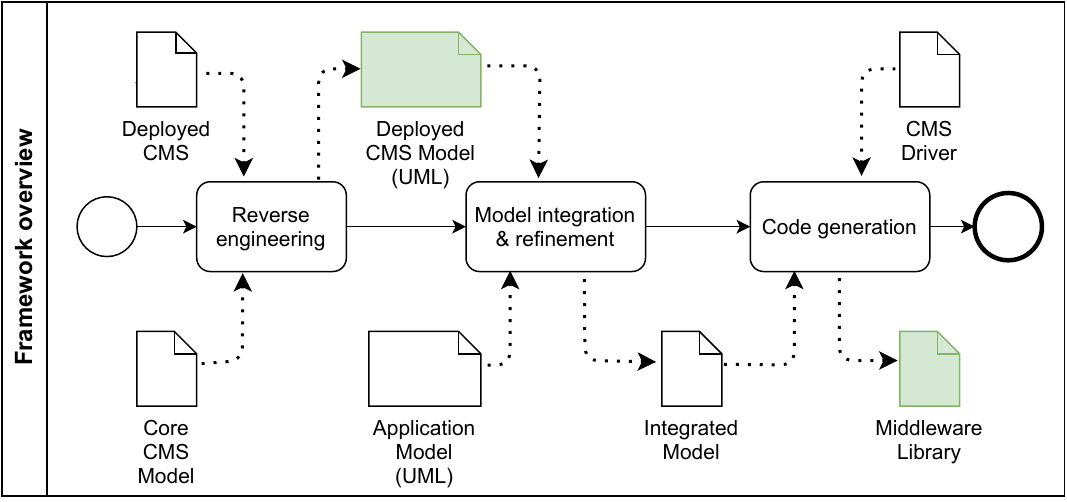}
    \caption{Framework overview}
    \label{fig:framework}
\end{figure}

In next sections we present in detail each of these steps.
\section{The \emph{Core CMS Model}}
\label{sec:CoreCMSModel}

\begin{figure}[t]
    \centering
    \includegraphics[width=0.9\columnwidth]{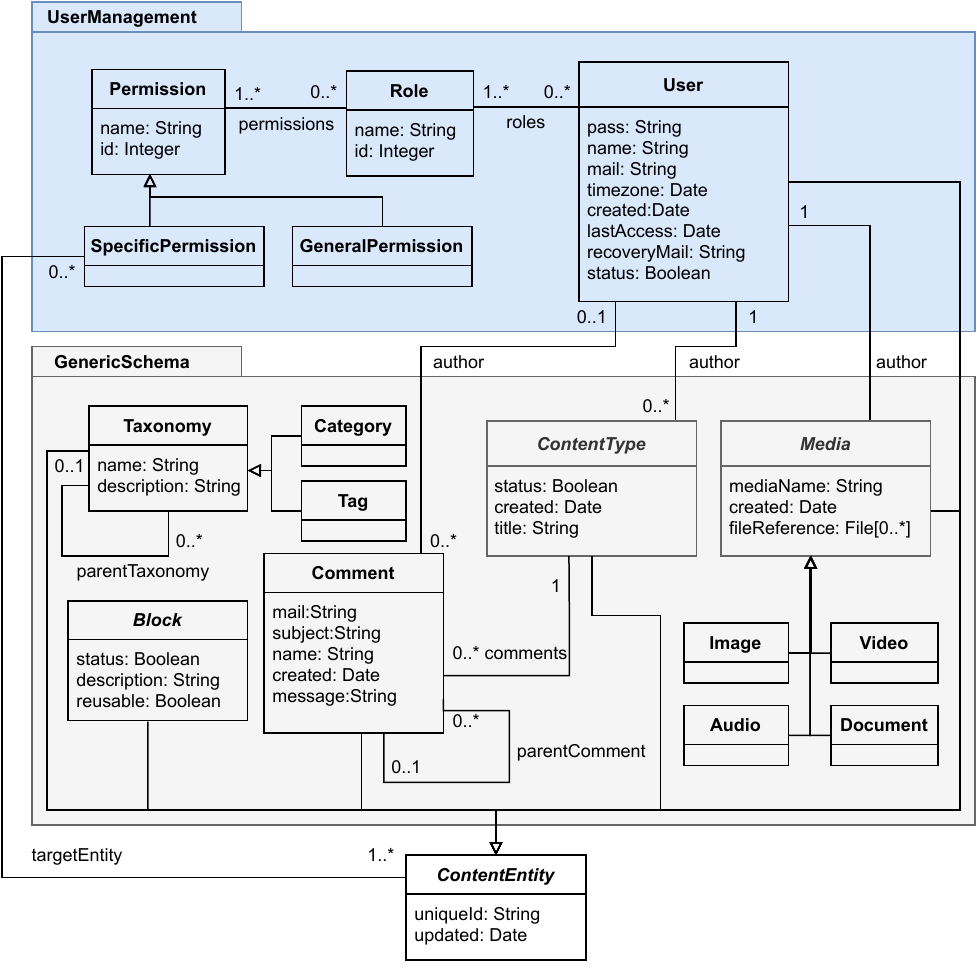}
    \caption{Schema and User Management of the Core CMS model}
    \label{fig:generic}
\end{figure}

This section presents our core model for CMS, representing the common concepts shared by all major CMS vendors.  The goal of this model is to facilitate the representation of the information schema behind specific CMSs by providing basic types that can be extended. This avoids having to start from scratch every time. Moreover the core elements also facilitate the management of CMS models when there is no need to access its detailed structure. This core model will be used by the reverse engineering process, see Section \ref{sec:reverse}, to create the complete model-based representation of an input CMS, 

We have come up with the elements in the core model after analyzing the three major CMS platforms (Drupal, WordPress and Joomla; as they together have over 70\% of the market share in this domain \cite{buildwith_2020,w3techs_2020}) and inferring the commonalities among them. Figure \ref{fig:generic} shows a partial view of this common model. We focus on the access control  and the content information parts. A more complete model, which also supports features such as revisions or translations of the content, is available in the online repository.

According to Figure \ref{fig:generic}, we can see that many classes inherit from \emph{ContentEntity}. This class provides the last update time of the entity and a unique identifier for it. This identifier will be used to query the CMS API and get the entity information and its linked resources.

Immediately below, we have \emph{ContentType}, that represents any content stored in the CMS. Typically, most CMSs offer at least pages and posts as content types, but any non-trivial CMS will come with several custom types to better represent the information domain the CMS is serving. Each custom type will be a subclass of this root \emph{ContentType} element. We can always associate a \emph{Comment} to a content type. Comments have a hierarchical relationship, allowing different threads of comments to coexist in parallel and can be associated with a user if the comment is done by a registered one. Content types can also be classified in taxonomies. Two typical taxonomies are \emph{Tags} and \emph{Categories}. As we have found these taxonomies in all CMSs we have directly included them in the core model but a specific CMS could also have custom taxonomies. Taxonomies also have a hierarchical relationship.

Moreover, CMSs also have \emph{Media} files that can be attached to the content, such as images, videos, or audio files. They are represented independently of the context where they are embedded, since media items can be reused across different content pieces. This also means that CMSs could also be used as a media repository. Finally, \emph{Blocks} represent complementary pieces of content of a \emph{ContentType}, for example, the ads attached to a particular content piece.

Beyond content management, CMS pay special attention to the access control and allow defining who can access the content following a standard Role-Based Access Control approach~\cite{mar:gar:cup:cup:cabot}, where every user has a set of \emph{Roles} attached, and every \emph{Role} has a set of permissions attached. \emph{Permission} represents a specific action that a user can do inside the system. There are two subtypes of permissions, the \emph{GeneralPermission} which is not attached to a particular domain entity, for instance, ``access to the admin dashboard'' and, the \emph{SpecificPermission} that is related to a specific content type, for example, ``edit video article''.

\section{The reverse engineering process}
\label{sec:reverse}

In this section, we explain the \emph{reverse engineering process}. This process is the first step in the integration framework and aims to extract the model of a deployed CMS. On one side, it starts by getting the URL and the user credentials of the target CMS to discover its API. On the other side, the process receives the \emph{Core CMS Model} presented above to perform a set of extensions over it to extract the model of the deployed CMS. As a result, we get the model in UML facilitating the integration with other parts of the system as shown in \mbox{Section \ref{sec:integration}}

In this section, we present a comparison of the API architecture between technologies, the discovery process followed to extract insights from the API, and the extension strategy followed over the \emph{Core CMS Model} to fit particular data scenarios.

\subsection{The CMSs API}

Concerning the API, the different CMSs implementations use a REST API to expose their content\footnote{Other API architectures as GraphQL are also implemented but not included in the standard installation. More information about the discussion can be found at \url{https://dri.es/headless-cms-rest-vs-jsonapi-vs-graphql}.}. These REST APIs follow different standards depending on the implementation. As an example, Drupal and Joomla follow JSON:API \cite{jsonapi_2020} as a way of structuring the answer and providing links to navigate between resources, while WordPress uses HAL \cite{hal_2020} as its standard way to provide linking between resources.

In contrast, all the APIs expose the schema of the CMS. As a schema, we mean the entities and attributes of the CMS and the custom ones created over the specific installation. Besides, if a change is done in the CMS schema, the API is updated automatically, meaning that we can consider this API as the current state of the CMS schema. Finally, all the analyzed technologies implement Open API\cite{openapi_2020} as a standard for API description in conjunction with custom discovery mechanisms specific to each technology. The \emph{Reverse Engineering} process uses the Open API descriptions and these custom methods to perform the discovery process.

\subsection{Discovery process}

To perform the extraction, we analyze the API by using the discovery methods of each CMS implementation. First, we systematically fetch all the API resources to detect all the entities exposed by the deployed installation. Then, we analyze the answer to infer from which type is every detected entity and which attributes has. Once we have spotted all the entities and their attributes, we analyze, again, the answers to detect their relationships.

\begin{lstlisting}[language=json,
label=lst:answerexample,
caption= Example excerpt of API answer, basicstyle=\ttfamily\scriptsize,]
"info": {
    "title": "Journal CMS",
},
"host": "example.com",
"basePath": "/api",
"definitions": {
    "node--videoarticle": {
        "title": "node:videoArticle Schmea",
        "description": "news main content",
        "properties": {
            "attributes": {
                "id": "Integer",
                "title": "String", 
                "...": "..."
            },
            "relationships": {
                "0": {
                    "type": "taxonomy--category",
                    "link": "..."
            }
        }
    }
}
\end{lstlisting}

In Listing \ref{lst:answerexample} we can see an example excerpt of Drupal's API answer. In this example we can see a \emph{node-{}-videoarticle} definition which is an entity that inherits from \emph{ContentType}. We guess the inheritance from the term "node" which refers to \emph{ContentType} in Drupal's terminology. Under \emph{properties} we can see a set of attributes as \emph{id}, \emph{title} and others (lines from 11 to 15). Finally, we can see that our entity has a relationship with a \emph{Taxonomy} of type \emph{Category} and the same answer provides us a link to fetch and discover this relationship (lines from 16 to 20).

If we detect entities that are not in the proposed \emph{core CMS} model, we can know from which types these entities inherit. For instance, in Listing~\ref{lst:answerexample}, \emph{VideoArticle} is a specific type of \emph{ContentType}, and therefore, we will create a new class extending our \emph{ContentType} class of the core model. This new class also will retain some specific annotations, such as the URL of the endpoint. It could be noted that this process is tied to every technology implementation as every technology builds the answer in different ways. Therefore, a specific extractor needs to be developed to support every CMS implementation.

\subsection{Extraction example}

\begin{figure}[tb]
    \centering
    \includegraphics[width=0.8\columnwidth]{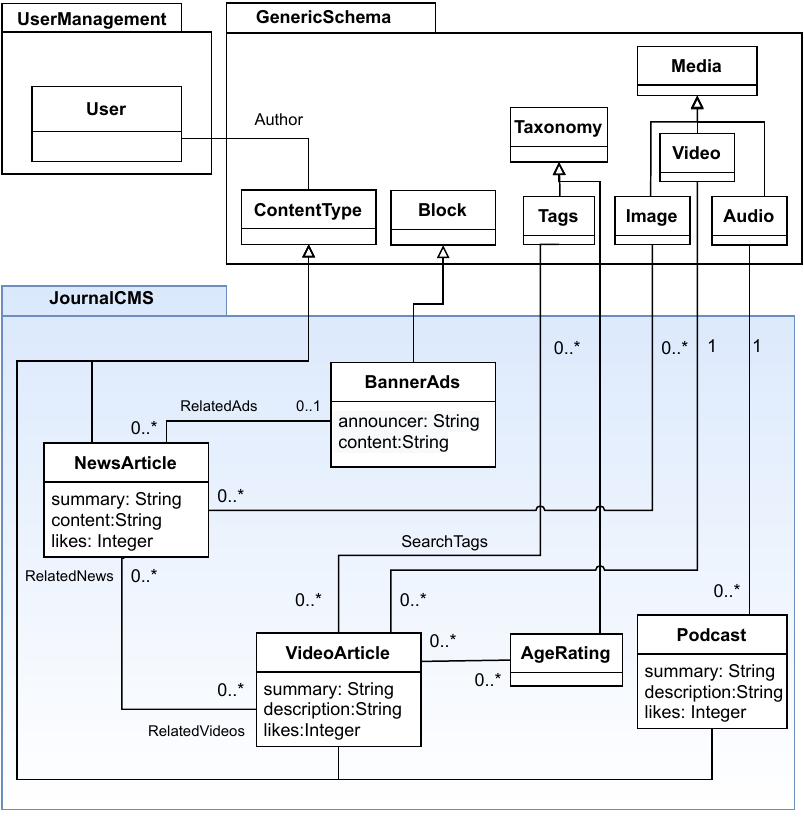}
    \caption{Example extracted model of a journal's CMS}
    \label{fig:extension}
\end{figure}

In Figure \ref{fig:extension} we have an example of an extracted model of a journal site. This example shows how the discovery process and the extensions over the \emph{Core CMS model} can describe complex data scenarios. This example describes a journal with different types of content such as video articles, written news, and podcasts with relationships between them.

In the figure, \emph{VideoArticle} represents the news edited as video. This class has a relationship with a \emph{Video} and some attributes, like \emph{likes}, which represent the number of likes made by users. In terms of Taxonomies, \emph{VideoArticle} have a relationship with \emph{AgeRating} which classifies the video by age recommendation. In addition, it also has a set of \emph{Tags} to facilitate the search of this content. Finally, this class has a relationship with a set of \emph{NewsArticle}. If a user wants further information about the news, this represents the related written news of the \emph{VideoArticle}.

\emph{NewsArticle} class, which extend from \emph{ContentType}, represents the written news. As previous, this class has attributes, like \emph{likes}, which represent the number of likes made by users. Also, the \emph{NewsArticle} has a relationship with \emph{BannerAds}, which extends from \emph{Block}, that represents the advertisements attached to a specific news. This relationship allows configuring the advertisements policy inside the CMS, and reproduce it also in the consumer apps. In addition, \emph{NewsArticle} can have a set of related \emph{VideoArticle} attached to it. In the journal's example, this represents written news that have their version as video news to be shown on a streaming platform or TV. Moreover, \emph{NewsArticle} also have media \emph{Images} as the images of news.

Finally, another main content of the site is \emph{Podcast}, which has a relationship with media \emph{Audio}, which represents content edited by audio. This content audio format has been gaining popularity in the last year and is a clear example of how CMS manages different types of content.

In conclusion, after the \emph{Reverse Engineering} process, we obtain the model of our deployed CMS in UML. This model can work with complex data scenarios, as we can see in Figure \ref{fig:extension}, and is ready to be integrated with other parts of the systems, as we see in the next section.

\section{Model integration and refinement}
\label{sec:integration}

The result of the reverse engineering process is a fully compliant UML class diagram. As such, we can use, combine, and refine that diagram as we would do with any other model of our information system. One of the most obvious integrations can be specifying how the user interface of different consumer components, such as web applications or mobile apps, query the extracted CMS model to retrieve the information they need to show to their users. 

We will use this scenario to continue with our running example. Let's assume our media company  is interested in building a mobile app to show a video news feed to its users.  More specifically, the app will show the video feed, let users \emph{like} some specific videos (and these \emph{likes} will be stored the CMS for future rankings and optimizations), access the details of a particular video and get related news to the ones she is watching. 

\begin{figure*}[t]
    \centering
     \includegraphics[width=0.7\columnwidth]{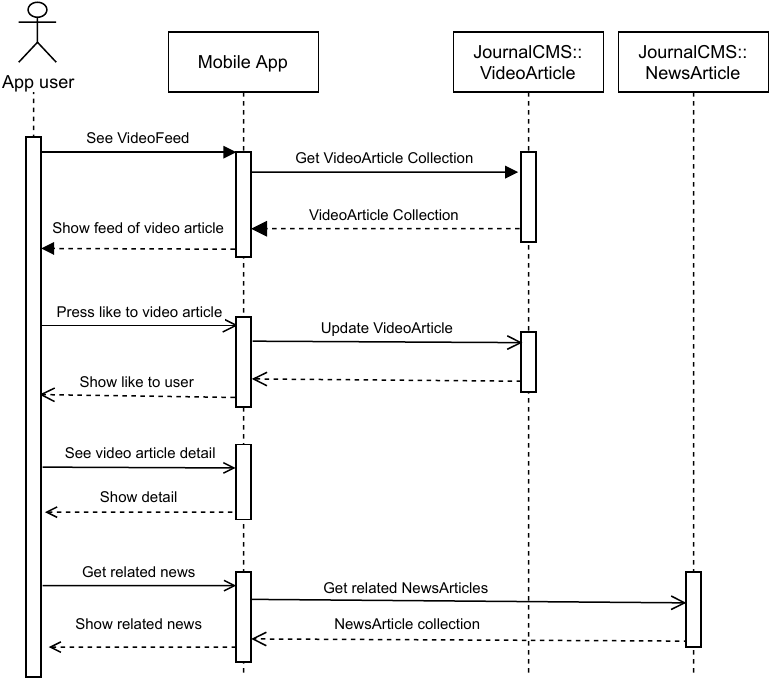}
    \caption{Integration example sequence}
    \label{fig:integrationSequence}
\end{figure*}

All this information is available in the Journal CMS. And since we now have it explicitly represented as a model, we just need to define the interaction between the UI app model and the CMS model. A possible sequence diagram is shown in Figure \ref{fig:integrationSequence}. In the diagram, when a user wants to see the video feed, a call to the \emph{VideoArticle} class is performed. When a user likes a specific video, this is translated to an update call to the journal's \emph{VideoArticle} class. Finally, when a user sees a video detail, the user can access the related content, querying, in that case, the journal's \emph{NewsArticle} class. 

The methods used in the sequence diagram are self-explanatory and, as we will see in the next section, will be made available in the next section via the code generation process. For more advanced scenario, we could also consider using a language like the Object Constraint Language \cite{richters2002} to more formally specify the exact queries to perform to the CMS.

Similarly, we could model other types of interactions or static diagrams with our CMS model. The key point is to realize that these interactions are modeled as any other type of interaction between two models and that when doing so the designer does not need to worry about the underlying CMS platform and version. The designer can stay at a platform-independent model. In fact, it would be possible to define a client consumer application that reads from two different CMS models without the designer not even being aware that the queried data elements belong to separate CMSs.

\section{Code generation infrastructure}
\label{sec:connector}

The final step of the framework is the \emph{code generation} process. This process aims to generate a middleware library to facilitate the communication between the deployed CMS and consumer applications. 

Regarding the presented integration example in Figure \ref{fig:integrationSequence}, the middleware would be in charge of the interaction between the \emph{Mobile app} and the content of the Journal's CMS such as \emph{VideoArticle} and  \emph{NewsArticle}. This middleware could work with other components (e.g., generating the mobile app) by providing the glue code required to ensure the communication between the app and the content source.

In this section, following Figure \ref{fig:generator}, we explain how this middleware is structured to be easily extensible for other technologies and how the consistent API is designed to abstract practitioners from the underlying technology. Finally, we provide a usage example of the middleware library following the integration example presented in Section \ref{sec:integration}.

\subsection{Middleware structure}

The middleware structure is composed of a set of generated classes from the extracted model, and a \emph{CMS Driver} composed by a set of static classes that allows us to move back and forward the data of the generated classes between the consumer app and the source CMS.

\begin{figure}[t]
    \centering
     \includegraphics[width=0.8\columnwidth]{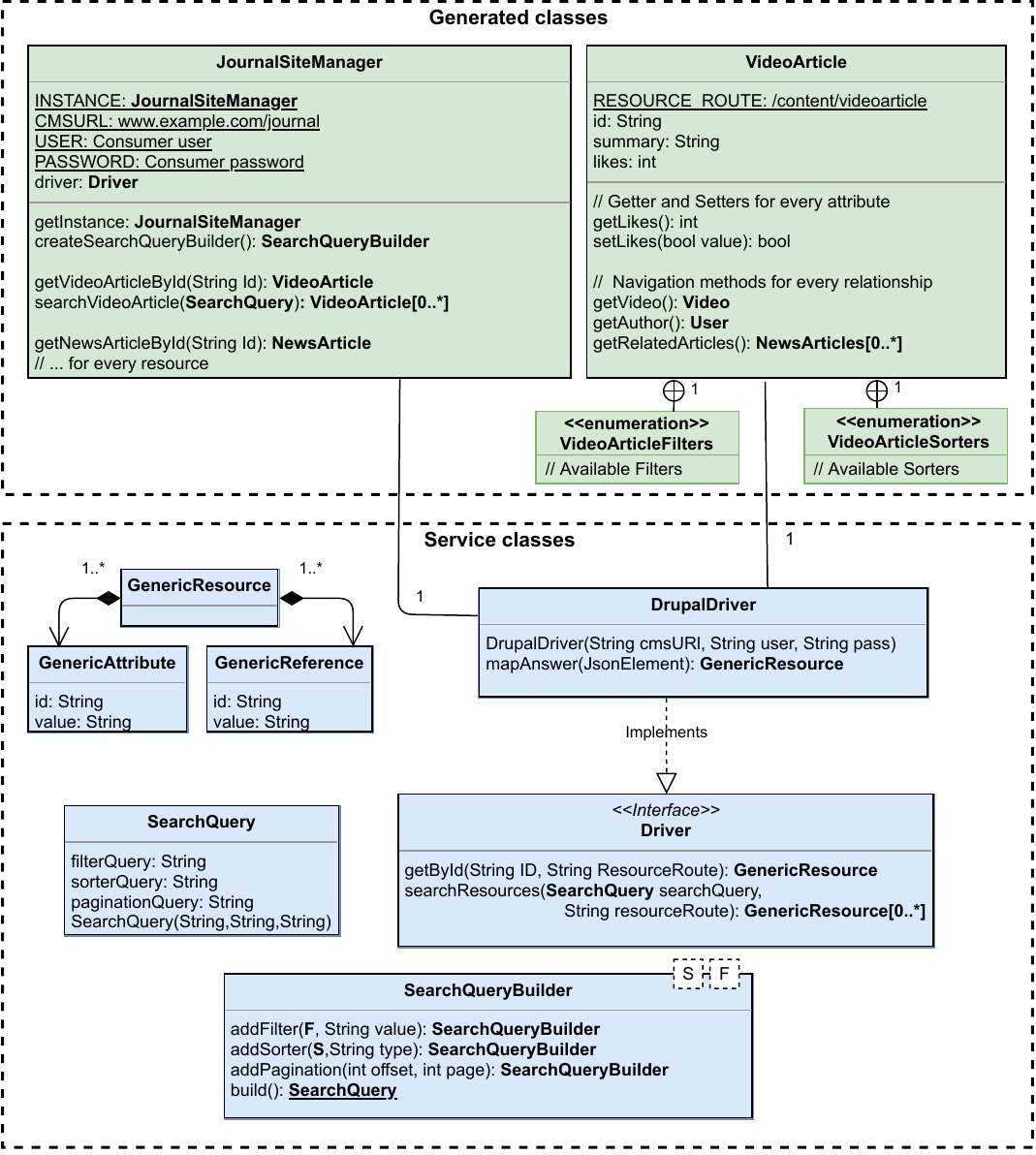}
    \caption{Middleware's implementation diagram: \emph{VideoArticle} example}
    \label{fig:generator}
\end{figure}

The generated classes are the dynamic part of the middleware. For each UML class of the extracted model, a java class is generated for it. This class contains the UML class attributes and relationships, and the possible sorter and filters to interact with the API. For instance, in Figure~\ref{fig:generator}, the class \emph{VideoArticle} corresponds to the class \emph{VideoArticle} of the example extracted model in Section~\ref{sec:reverse}. In addition to the previous ones, a \emph{JournalSiteManager} is generated. This class is built using the singleton pattern and represents the whole CMS site, and brings access to the generated classes with specific methods for each one.

The \emph{CMS Driver} classes are the static part of the middleware. The \emph{Driver} class is in charge of the communication with the source CMS and is the only one tied to a particular technology. This approach means that the driver is the only class we need to reimplement to extend support for new CMS technologies. This driver has a single relationship with the \emph{JournalSiteManager} class. In Figure \ref{fig:generator}, we can see a \emph{DrupalDriver} class as our journal's site is powered by Drupal.

Also, inside the \emph{CMS Driver}, and to ensure the separation between the agnostic part and the specific part, we propose the \emph{GenericResource} class as an interface between the generated classes and the driver. In addition, we suggest a \emph{SearchQueryBuilder} class as a way to provide a consistent API to practitioners in terms of fetching the source CMS. At last, \emph{SearchQuery} is the result of the query builder being an immutable class and allowing the \emph{DrupalDriver} class to build the particular final query to the CMS API.

In summary, the middleware is composed of a set of dynamic classes generated from the extracted model and a set of already developed static classes representing the \emph{CMS Driver}. 

\subsection{The middleware API}

One of the key points is to facilitate the usage of this middleware without regarding the underlying technology. To do so, we have proposed a consistent API as a standard method to interact with them. We have inspired the design of the consistent API in the identified shared concepts of the different open-source CMS APIs.

The main methods to interact with the content are present in the \emph{JournalSiteManager} of our example. This class has a \emph{getById} and a \emph{search} methods for every generated class from the extracted model. The first one needs a content id to return the result, and the second one needs a \emph{SearchQuery} to do so. 

The \emph{SearchQueryBuilder} is a query builder that wraps up the concepts of filtering, ordering, and pagination mentioned before. It provides a way of building queries benefiting from the main API features without regarding its underlying implementation. Therefore, we can add filters, sorters, and pagination using the provided methods. Once the search query is built, it returns a \emph{SearchQuery} as an immutable class.

Finally, one of the key points of the analyzed REST API is hypermedia. Hypermedia is the ability to navigate through resources to discover the API. For example, if we get a \emph{VideoArticle} and we want to get a specific related \emph{NewsArticle}, we know from the answer itself how to retrieve this related content. Instead of automatically getting the related content, we have decided to implement this navigation method to avoid recursive calls. In Figure \ref{fig:generator} we can see an example of this in the \emph{VideoArticle} class that has navigation methods. From a \emph{VideoArticle} we can get the related \emph{NewsArticle} or the \emph{User} corresponding to the author. 

\subsection{Middleware usage example}

In Listing \ref{lst:classexample}, we can see an example of the \emph{Mobile App} class of Figure \ref{fig:integrationSequence} using the generated middleware. In this example, we can see how practitioners could easily integrate the Journal CMS as a content source without regarding the underlying technology. This example uses the java code generated by the tool presented in the next section following the integration example of Section \ref{sec:integration}.

\begin{lstlisting}[
float=!b,
firstnumber=1,
label=lst:classexample,
language=java,
basicstyle=\ttfamily\scriptsize,
caption= Example MobileApp class demonstrating the use of the middleware library,
numbers=left,
xleftmargin=5.0ex]
class MobileApp() {

  public void main(){
    // 1 - Get the video feed
    List<VideoArticle> feed = getVideoFeed();
    // 2 - User press like to the first video
    VideoArticle firstVideo = feed.get(0);
    firstVideo.setLike(true);
    // 3 - User wants to see the related news of the video
    List<NewsArticle> relatedNews = firstVideo.getRelatedArticles();
  }
  public List<VideoArticle> getVideoFeed() {
    // The site manager class
    JournalSiteManager site = JournalSiteManager.getInstance();
    // We get a search query builder
    SearchQueryBuilder queryBuilder = site.getSearchQueryBuilder();
    // We call the search method with an empty query
    return site.searchVideoArticle(queryBuilder.build());
  }
}
\end{lstlisting}

In the main method, we perform the sequence represented in Figure \ref{fig:integrationSequence}. Therefore, in line 5, we get the video feed by calling the \emph{getVideoFeed()} method. In this method, we get an instance of \emph{JournalSiteManager} class as it provides methods to interact with all the generated classes. Then, to get the list of videos we call the method \emph{searchVideoArticle()} form the \emph{JournalSiteManager} in line 18. This method requires a \emph{SearchQuery} as an entry parameter. In our case, we get a \emph{searchQueryBuiler}, but we do not need to set any filter or sorter configuration. So we build it directly in line 18. The method returns a list of \emph{VideoArticles}.

Then, a user presses the \emph{like} button of the first video of our video feed. Therefore we get the pressed video in line 7, and we call the \emph{setlike()} method of it in the next line. Finally, in line 10, to retrieve the related content, we use the navigation methods of the \emph{VideoArticle} class calling the \emph{getRelatedNews()} method. This methods returns the list of \emph{NewsArticles} related to the particular \emph{ViodeArticle}.

\section{Tool support}
\label{sec:tool}

We have developed a prototype implementation of our framework, available under the Eclipse Public License\footnote{\url{https://www.eclipse.org/legal/epl-2.0}}. The source code of the project and the Eclipse update site are available in the public repository.

Our implementation is built in Java and relies on the Eclipse Modeling Framework~\cite{web:emf} for the core modeling support. In particular, we have used Xtext and Xtend~\cite{web:xtend} to implement the code-generation process. WordPress and Drupal are the two CMSs supported by our tool, but others could be easily integrated. Further instructions about the usage of this tool are provided on the public repository.

\section{Related Work}
\label{sec:relwork}

Several research works have investigated the intersection between CMS and MDE. A first group of works have focused on modeling specific CMS components or extensions/plugins. Priefer \cite{prief:wolf:stru} proposes a model-driven approach to generate Joomla extensions. Martínez \cite{mar:gar:cup:cup:cabot} presents a metamodel for the access-control rules in a CMS to be used for verification and validation purposes. Our work targets the complete CMS definition and not just specific components.

Another group of works focus on the generation of a new CMS implementation. Trías~\cite{trias:cast:lop:marc} propose a CMS common metamodel to capture the key concerns required to model CMS-based web applications and, Qafmolla \cite{qafmolla:cuon:richta} propose the generation of CMS instances from it. Similarly, Souer \cite{soue:kuper:hemls} aims to generate a CMS from an automatically generated configuration description created by business users. Modeling languages to define new CMS is also the goal of Saraiva \cite{sar:sil:rod}. Our work is different from these approaches in that: 1) Our model is not generic but extensible to precisely represent the information schema of the CMS (and not just global concepts as \textit{post} or \textit{page}) and 2) our target is the integration of the CMS with the rest of the system and therefore our framework comprises the interactions between the data consumer components and the CMS and 3) we start from an exiting CMS, which, in our opinion, is a more realistic scenario as most companies already have a CMS in place and are not looking to start from scratch but evolve it. 

It is also worth mentioning works that have similar goals to ours, but that target different information sources. SQL databases (e.g. Egea \cite{egea2010mysql4ocl}, and Nguyen \cite{Nguyen2019}), NoSQL databases (e.g. \cite{daniel2016umltographdb}, \cite{abdelhedi2017}, \cite{comyn2017model}), Data Warehouses (e.g. \cite{trujillo2003uml}) and even spreadsheets \cite{querysheet} have been integrated in MDE processes.

Finally, given that headless CMSs expose their data through a REST API, we comment on model-based proposals targeting REST APIs. Some relevant examples are \cite{hamza:izquierdo:cabot}, presenting an MDE-based tool to create, visualize, manage, and generate OpenAPI definitions, \cite{izquierdo:cabot}, proposing an MDE process to discover and composite JSON documents or \cite{rivero_extensible_2014}, focusing on building REST APIs directly from user requirements. In our case, we do not aim to generate new APIs but to provide a specific reverse engineering process adapted to the concrete REST API endpoints and patterns offered by popular CMSs. In this sense, some concepts of these works have been integrated in our approach as building blocks on top of which we have built the specific connectors we required for the CMSs.

\section{Conclusions and future work}
\label{sec:conclusions}

In this work, we have proposed a model-based framework to integrate headless CMSs in a development process. Thanks to our framework, the conceptual schema behind the CMS can be easily exposed and reused as part of the global development process. The framework comprises a core CMS model, a reverse engineering process that extends it to precisely represent the data structure of an input CMS, and a middleware that enables modeling front-ends that need to consume that data in a platform-agnostic way. 

This facilitates the reusability and evolution of the software system, also enabling potential migrations among different CMS versions and platforms, e.g., to benefit for more advanced features or scalability improvements, while minimizing the impact of such changes on all the deployed clients (e.g. mobile apps).

As further work, we will study how the reverse engineering process could take into account presentation suggestions that are now also being embedded in the CMS headless API to achieve a more homogeneous look\&feel across different client front-ends. There are some important trade-offs here that deserve to be analyzed. We would also like to explore the benefits of  exposing CMS data in other domains, like Machine Learning, where easy access to CMS data could be exploited for training ML models. In this context, deploying our framework as a plugin for model-based data science tools like KNIME\cite{knime} could be interesting. Finally, at the tool level, we plan to extend our support to other CMSs and to validate the tool in industrial settings.
\newpage
\section*{Acknowledgements}

The research described in this paper has been partially supported by the AIDOaRT Project, which has received funding from the ECSEL Joint Undertaking (JU) under grant agreement No 101007350. The JU receives support from the European Union’s Horizon 2020 research and innovation programme and Sweden, Austria, Czech Republic, Finland, France, Italy, and Spain. 

In addition, the research has also been partially supported by the TRANSACT project, which has received funding from the ECSEL Joint Undertaking (JU) under grant agreement No 101007260. The JU receives support from the European Union's Horizon 2020 research and innovation programme and Netherlands, Finland, Germany, Poland, Austria, Spain, Belgium. Denmark, Norway.

%
%
%

\end{document}